\documentclass{article}\usepackage{jkas2}
\newcommand{\be}{\begin{equation}}
\newcommand{\ee}{\end{equation}}
\newcommand{\ba}{\begin{eqnarray}}
\newcommand{\ea}{\end{eqnarray}}

\runningauthor{Lazarian}
\runningtitle{Turbulence from Observations}
\beginpage{1}
\endpage{5}

\begin{document}

\font\twelvei = cmmi10 scaled\magstep1 
       \font\teni = cmmi10 \font\seveni = cmmi7
\font\mbf = cmmib10 scaled\magstep1
       \font\mbfs = cmmib10 \font\mbfss = cmmib10 scaled 833
\font\msybf = cmbsy10 scaled\magstep1
       \font\msybfs = cmbsy10 \font\msybfss = cmbsy10 scaled 833
\textfont1 = \twelvei
       \scriptfont1 = \twelvei \scriptscriptfont1 = \teni
       \def\mit{\fam1 }
\textfont9 = \mbf
       \scriptfont9 = \mbfs \scriptscriptfont9 = \mbfss
       \def\bmit{\fam9 }
\textfont10 = \msybf
       \scriptfont10 = \msybfs \scriptscriptfont10 = \msybfss
       \def\bmsy{\fam10 }

\def\etal{{\it et al.~}}
\def\eg{{\it e.g.,~}}
\def\ie{{\it i.e.,~}}
\def\lsim{\raise0.3ex\hbox{$<$}\kern-0.75em{\lower0.65ex\hbox{$\sim$}}}
\def\gsim{\raise0.3ex\hbox{$>$}\kern-0.75em{\lower0.65ex\hbox{$\sim$}}}

\title{Turbulence Statistics from Spectral Line Observations}

\author{A. Lazarian}
\address{Department of Astronomy, University of Wisconsin-Madison\\
{\it E-mail: lazarian@astro.wisc.edu}}

%


\address{\normalsize{\it ()}}

\abstract{Turbulence is a crucial component of dynamics of astrophysical
fluids dynamics, including those of ISM, clusters of galaxies and 
circumstellar regions. 
Doppler shifted spectral lines provide a unique source of   
information on turbulent velocities. We discuss Velocity-Channel
Analysis (VCA) and its offspring Velocity Coordinate Spectrum
(VCS) that are based on the analytical description of the
spectral line statistics.
Those techniques are well suited for studies of supersonic
turbulence. We stress that a
 great advantage of VCS is that it does not necessary
require good spatial resolution.
Addressing the studies of mildly supersonic and subsonic
turbulence we discuss the criterion
that allows to determine whether Velocity Centroids are
dominated by density or velocity. We briefly discuss 
ways of going beyond power spectra by  using
of higher
order correlations as well as genus analysis. 
We outline the relation between Spectral Correlation
Functions and the statistics available through VCA and VCS.
}

\keywords{turbulence, molecular clouds, MHD}

\maketitle

\section {Why should we study astrophysical turbulence?}

As a rule astrophysical fluids are turbulent and the turbulence 
is magnetized. This ubiquitous turbulence determines the transport
properties of interstellar medium (see 
Elmegreen \& Falgarone 1996, Stutzki 2001,
Cho et al. 2003) and intracluster medium
(Inogamov \& Sunyaev 2003, Sunyaev, Norman \& Bryan 2003,
see review by Lazarian \& Cho 2004a), many properties of Solar and stellar
winds, accretion disks etc. One may say that to understand heat 
conduction, transport and acceleration of cosmic rays, propagation
of electromagnetic radiation in different astrophysical environments it
is absolutely essential to understand the properties of underlying turbulence.
The mysterious processes of star formation (see McKee \& Tan 2002, 
Elmegreen 2002, Pudritz 2001) and interstellar chemistry (see Falgarone 
1999 and
references therein), shattering and
coagulation of dust (see Lazarian \& Yan 2003 and references therein) 
are also intimately related
to properties of magnetized compressible turbulence (see reviews by
Elmegreen \& Scalo 2004).

From the point
of view of fluid mechanics astrophysical turbulence 
is characterized by huge Reynolds numbers, $Re$, which is  
the inverse ratio of the
eddy turnover time of a parcel of gas to the time required for viscous
forces to slow it appreciably. For $Re\gg 100$ we expect gas to be
turbulent and this is exactly what we observe in HI (for HI $Re\sim 10^8$).

Statistical description is a nearly indispensable strategy when
dealing with turbulence. The big advantage of statistical techniques
is that they extract underlying regularities of the flow and reject
incidental details. Kolmogorov description of unmagnetized incompressible
turbulence is a statistical
one. For instance it predicts that the difference in velocities at
different points in turbulent fluid increases on average
with the separation between points as a cubic root of the separation,
i.e. $|\delta v| \sim l^{1/3}$. In terms of direction-averaged
energy spectrum this gives the famous Kolmogorov
scaling $E(k)\sim 4\pi k^2 P({\bf k})\sim k^{5/3}$, where $P({\bf k})$ 
is a {\it 3D} energy spectrum defined as the Fourier transform of the
correlation function of velocity fluctuations $\xi ({\bf r})=\langle  
\delta v({\bf x})\delta v({\bf x}+{\bf r})\rangle$. Note that in
this paper we use $\langle  ...\rangle$ to denote averaging procedure.

The example above shows the advantages of the statistical approach
to turbulence. For instance, the energy spectrum 
$E(k)dk$ characterizes how much
energy resides at the interval of scales $k, k+dk$. At large scales $l$
which correspond to small wavenumbers $k$ ( i.e. $l\sim 1/k$) one expects
to observe features reflecting energy injection. At small scales
one should see the scales corresponding to
sinks of energy. In general, the shape of the spectrum is
determined by a complex process of non-linear energy transfer and
dissipation. For Kolmogorov turbulence the spectrum 
over the inertial range, i.e. the range where neither  energy injection nor
energy dissipation are important, is
characterized by a single power law and therefore self-similar.
Other types of turbulence, i.e. the turbulence of non-linear waves
or  the turbulence of shocks, are characterized by different power
laws and therefore can be distinguished from the Kolmogorov turbulence
of incompressible eddies.    
  
In view of the above it is not surprising that attempts
to obtain spectra of interstellar turbulence have been numerous. In fact they
date as far back as the 1950s
(see von Horner 1951, Munch 1958,
Wilson et al. 1959). However, various directions
of research achieved various degree of success (see 
Kaplan \& Pickelner 1970, a review by Armstrong, Rickett
\& Spangler 1995). 
For instance, studies of turbulence statistics of ionized media 
were more successful
(see Spangler \& Gwinn 1990) and provided the information of
the statistics of plasma density at scales $10^{8}$-$10^{15}$~cm. 
This research profited
a lot from clear understanding of processes of scintillations and scattering
achieved by theorists (see Narayan \& Goodman 
1989). At the same time 
the intrinsic limitations of the scincillations technique
are due to the limited number of sampling directions, and  relevance only to
ionized gas at extremely small scales.
Moreover, these sort of measurements provide only the density statistics, 
which is an indirect measure of turbulence. 

Velocity statistics is much more coveted turbulence measure. 
Although, it is clear that Doppler broadened lines
are affected by turbulence, recovering of velocity statistics was
extremely challenging without an adequate theoretical insight.
Indeed, both velocity
and density contribute to fluctuations of the intensity in the 
Position-Position-Velocity (PPV) space.  

In this review we concentrate on discussing recent progress in analytical
description of the PPV statistics. We mostly discuss power spectra, but
also try to look beyond it. Some of the interesting tools, like wavelets or Principal
Component Analysis are beyond the scope of this review and will be discussed
elsewhere.

\section {How good are Velocity Centroids?}

Let us  consider ``unnormalized'' velocity centroids:
\begin{equation}
S(\mathbf{X})=\int v_{z}\ \rho_{s}(\mathbf{X},v_{z})\ {\rm d}v_{z},
\label{eq:S}
\end{equation}
where $\rho_{s}$ is the density of emitters in the  PPV space\footnote{ Traditionally
used centroids include normalization by the integral of $\rho_s$. This,
however does not substantially improve the statistics, but makes the
analytical treatment very involved (Lazarian \& Esquivel 2003).}.

In the case of emissivity proportional to the first power of density and
provided that the turbulent region is thin for its radiation,
analytical expressions for structure functions\footnote{Expressions for the correlation 
functions are straightforwardly related to those of structure functions (Monin \& Yaglom 1975). 
The statistics of centroids using correlation functions was used in a later paper by 
Levier (2004).}   of centroids, i.e.
$ 
\left\langle \left[S(\mathbf{X_{1}})-S(\mathbf{X_{2}})\right]^{2}\right\rangle $
 were derived in Lazarian \& Esquivel (2003, henceforth LE03). In that paper the following criterion
for centroids to reflect the statistics of velocity\footnote{LE03 showed that the 
solenoidal component of 
velocity spectral tensor can be obtained from observations
using velocity centroids in this regime.} was established:
\begin{equation}
\left\langle \left[S({\bf X_1})-S({\bf X_2})\right]^2 \right\rangle \gg
\langle V^2 \rangle \left\langle \left[I({\bf X_1})-
I({\bf X_2})\right]^2 \right\rangle,
\label{criterion}
\end{equation}
where $I({\bf X})$ is the intensity $I({\bf X})\equiv \int \rho_s dV$ and 
the velocity dispersion $\langle V^2 \rangle$ can
be obtained using the second moment of the spectral lines:
\begin{equation}
\langle V^2 \rangle\equiv \langle \int_{\bf X} V^2 \rho_s dV\rangle/
\langle \int_{\bf X} \rho_s dV
\rangle .
\label{V2}
\end{equation}

LE03 proposed to subtract the right hand sight of the expression 
(\ref{criterion})
from the left hand sight of (\ref{criterion}) to obtain {\it modified velocity centroids}
that may still reflect velocity statistics 
even when ordinary centroids are dominated
by density contribution. A numerical study in Esquivel \& Lazarian (2005) confirmed that
 the criterion given by eq~(\ref{criterion}) is correct and revealed that for MHD turbulence
simulations it holds for turbulence with Mach number less than 2. This was consistent with
an earlier analysis of a different set of MHD turbulence data by Brunt \& Mac
Low (2003) who observed 
that velocity centroids poorly present velocity statistics for high Mach number turbulence.
Esquivel \& Lazarian (2005) studied modified velocity centroids and
the concluded that the velocity centroids was
an OK technique to study subsonic turbulence as well as very mildly supersonic turbulence.
The modified centroids are better suited for obtaining velocity statistics as the
Mach number increased. In addition the aforementioned study has proven using MHD simulations
data that an analytical approach employed in LE03 provides a good representation of
centroids.

\section {What is the Velocity Channel Analysis (VCA)?}

Power spectra of fluctuations measured within narrow ranges of velocities
have been observed by different authors at various times (see
Green 1993). Those power spectra were guessed to be related
to underlying turbulence, but what exactly those observed spectra
mean was completely unclear.

This problem was addressed in Lazarian \& Pogosyan (2000, henceforth LP00)
who found the relation between the  the spectrum of
intensity fluctuations in channel maps and underlying spectra of velocity and density.
They found that the power index of the intensity fluctuations depends on the
thickness of the velocity channel (see Table~1).

\begin{table*}
\caption{\label{t:lazarian+pogosyan}
A summary of analytical results for channel map statistics derived in LP00.}
\begin{tabular}{lcc}
\noalign{\smallskip} \hline \noalign{\smallskip}
Slice & Shallow 3-D density & Steep 3-D density\\
thickness & $P_{n} \propto k^{n}$, $n>-3$&$P_{n} \propto k^{n}$, $n<-3$\\
\noalign{\smallskip} \hline \noalign{\smallskip}
2-D intensity spectrum for thin~~slice &
$\propto K^{n+m/2}$    & $\propto
K^{-3+m/2}$   \\
2-D intensity spectrum for thick~~slice & $\propto K^{n}$
& $\propto K^{-3-m/2}$  \\
2-D intensity spectrum for very thick~~slice & $\propto K^{n}$ & $\propto \
K^{n}$  \\
\noalign{\smallskip} \hline \noalign{\smallskip}
\end{tabular}

{{\it thin} means that the channel width $<$ velocity dispersion at the scale under
study}\\
{{\it thick} means that the 
channel width $>$ velocity dispersion at the scale under
study}\\
{{\it very thick} means that a
substantial part of the velocity profile is integrated over}\\
{$m$ is the power-law index of velocity structure
function, i.e. $\langle (v(x+r)-v(x))^2\rangle \sim r^{m}$.}\\
{$K$ is a 2D wavevector in the plane of the slice, $k$ a 3D wavevector.}
\end{table*}

It is easy to see that both steep and shallow underlying density
the power law index
{\it steepens} with the increase of velocity slice
thickness. In the thickest velocity slices the velocity information
is averaged out and it is natural that we get the
density spectral index $n$. The velocity fluctuations dominate in
thin slices, 
and the index $m$ that characterizes the velocity  fluctuations,
i.e. $|\delta v|\sim l^{m/2}$,
can be obtained using thin velocity slices (see Table~1). Note, that the
notion of thin and thick slices depends on a turbulence scale under
study and the same slice can be thick for small scale turbulent fluctuations
and thin for large scale ones. The formal criterion for the slice to be
thick is that {\it the dispersion of turbulent velocities on the scale studied
should be less than the velocity slice thickness}.  Otherwise
the slice is {\it thin}.

One may notice that the spectrum
of intensity in a thin slice gets shallower as the underlying
velocity get steeper. To understand this effect let us consider turbulence
in incompressible optically thin media. The intensity of emission
in a slice of data is proportional to the number of atoms per
the velocity interval given by the thickness of the data slice.
Thin slice means that the velocity dispersion at the scale of
study is larger than the thickness of a slice. The increase
of the velocity dispersion at a particular scales means that
less and less energy is being emitted within the velocity
interval that defines the slice. As the result the image of
the eddy gets fainter. In other words, the larger is the
dispersion at the scale of the study the less intensity
is registered at this scale within the thin slice of spectral
data. This means that steep velocity spectra that correspond
to the flow with more energy at large scales should produce
intensity distribution within thin slice for which the
more brightness will be at small scales. This is exactly
what our formulae predict for thin slices (see also LP00).

The result above gets obvious when one recalls that the largest
intensities within thin slices are expected from the regions that
are the least perturbed by velocities. If density variations are
also present they modify the result. When the amplitude of density
perturbation becomes larger than the mean density, both the
statistics of density and velocity are imprinted in thin slices.
For small scale asymptotics of thin slices this happens, however, only when the density spectrum
is shallow, i.e. dominated by fluctuations at small scales.

Predictions of LP00 were tested in Lazarian et al. (2001) and Esquivel et al. (2003)
using numerical MHD simulations of compressible interstellar gas.
Simulated data cubes allowed both density and velocity statistics
to be measured directly. Then these data cubes were
used to produce synthetic spectra which were analyzed using the
VCA. As the result, the velocity
and density statistics were successfully recovered.
Thus one can confidently apply  the
technique to HI observations. Note, that the VCA 
is only sensitive to velocity\footnote{For Kolmogorov turbulence the
velocity gradient scales as $l^{-2/3}$.}  
and density gradients on the scale under 
study, so the regular Galactic rotation curve or large scale 
distribution of emitting gas does not interfere with the analysis.
This was tested in Esquivel et al. (2003). The main formulae of LP00
were also obtained  also in Lazarian \& Pogosyan (2004) as the limiting
case of zero absorption within a more general treatment.

\section {What is Velocity Coordinate Spectrum (VCS)?}

Spatial resolution is essential for centroids and channel maps. However,
the velocity fluctuations are also imprinted on the 
fluctuations of intensity along the velocity coordinate direction. 
The corresponding 3D PPV spectra
were derived in LP00. Table~3 in LP00 states that two terms, one
depending only on velocity and the other depending both on velocity
and density, contribute to the spectrum measured along velocity
coordinate. 
for steep density the intermediate scaling therefore is
$k_z^{2n/m}$ ($n$ is negative), while the small scale assymptotics scales as 
$k_z^{-6/m}$. If the density is shallow the situation is reversed,
namely, at larger scales $k_z^{-6/m}$ assymptotics dominates, while
$k_z^{2n/m}$ assymptotics is present at smaller scales. The transition
from one assymptotics to another depends on the amplitude of density
fluctuations. Therefore both density and velocity statistics can 
be restored from the observations this way. If the measurements are done
with an instrument of poor spatial resolution these are the
expected scalings. A further study of these
interesting regime is done in Chepurnov \& Lazarian (2004), where
in addition to the parallel lines of sight, that is a default for
LP00 and LP04 studies, the converging lines of sight are taken into
account. The latter case is typical for studies of HI at high galactic
latitudes. 
The requirements for the VCS to operate is for the turbulence 
to be
{\it supersonic}, for the instrument to have an adequate {\it
spectral} resolution, and the {\it signal to noise ratio} to
be high. The latter requirement is due to the necessity of
measuring steep spectra along velocity coordinate (in $z$-direction)
 when the resolution
is low.

When the fluctuations are resolved, the velocity coordinate spectrum
is more shallow, namely, $k_z^{2(n+2)/m}$ and $k_z^{-2/m}$ (see Table~5
in LP00). As in the previous case, the shallow of the spectra, which
depends
on the velocity and density indexes $m$ and $n$, dominates at the 
smaller scales. For shallow spectra of intensity the requirement for
the signal to noise is less stringent. The transition from one
regime to another is  shown in Fig.~1.

\begin{figure} [h!t] 
{\centering \leavevmode 
\epsfxsize=2.5in\epsfbox{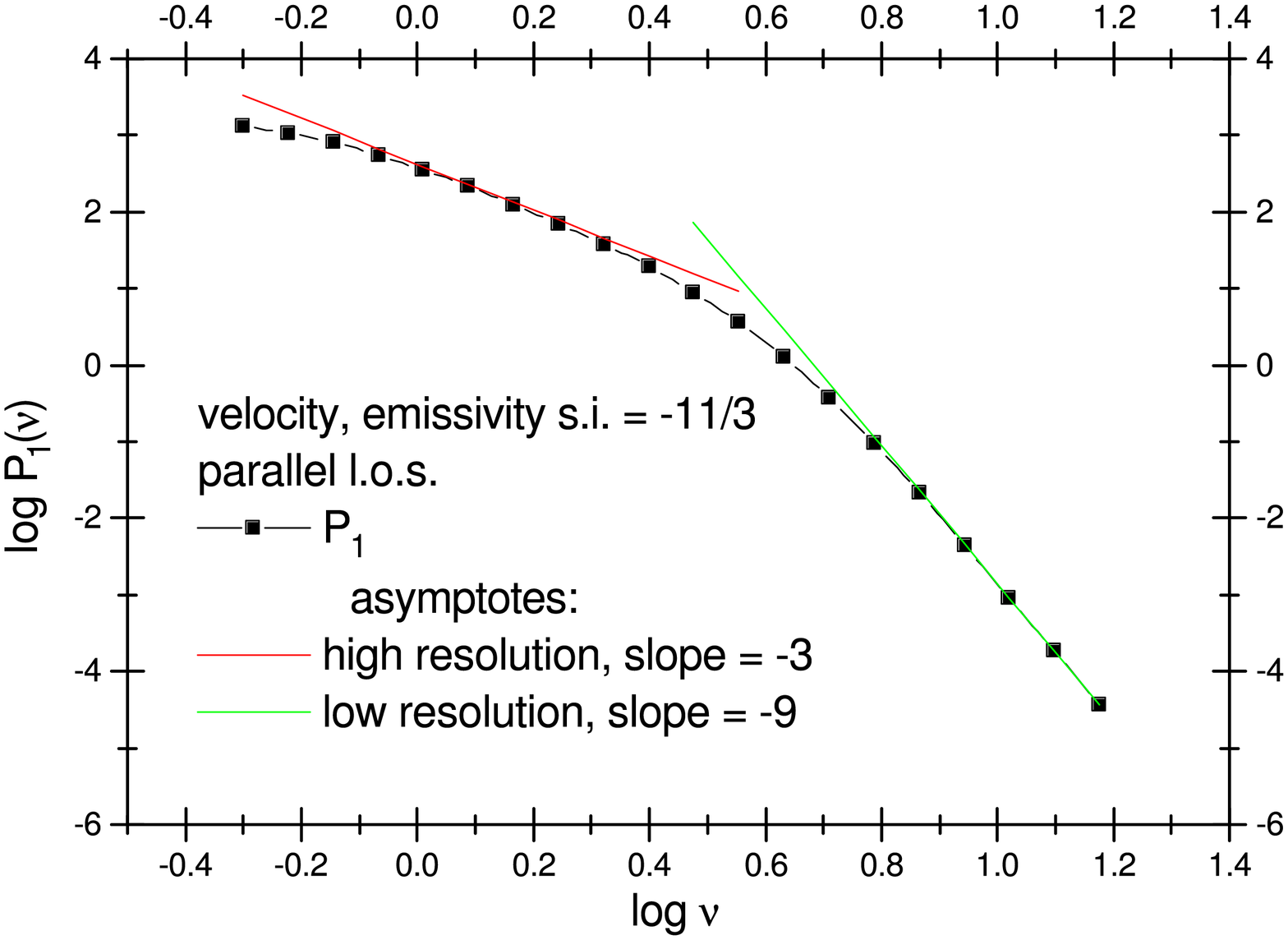} 
} 
\caption{One dimensional power spectrum along the velocity coordinate by
Chepurnov \& Lazarian (2004). As the beam size gets larger than the 
size of the eddies under study, the spectral index undergoes a change
between the two assymptotics derived in LP00. Only the spectrum
arising from velocity fluctuations is shown. Note that $k_z$ discussed in
the text corresponds to frequency $\nu$ shown in the figure.}  
\end{figure}

For the crossing lines of sight geometry the spectrum found in Chepurnov
\& Lazarian (2005) is not so steep as  in the case of parallel lines
of sight. Therefore the problem of signal to noise is mitigated  for
such a study.

Potentially VCS provides a unique technique for studies turbulence in
extragalactic objects where the spatial resolution is poor. So far,
VCS has been successfully tested with synthetic maps only. This new
technique should be applied to the existing and new spectral line datasets.

\section {What is the effect of Absorption?}

The issues of absorption were worrisome for the researchers from the
very start of the research in the field (see Munch 1958). The erroneous
statements about the effects of absorption on the observed turbulence
statistics are widely spread in the literature. For instance, a
fallacy that absorption allows to observe density fluctuations
localized in the thin surface layers of clouds, i.e. 2D turbulence,
exists (see discussion in LP04).

Using transitions that are less affected by absorption, e.g. HI,
allows frequently to avoid the problem. However, it looks foolish
to disregards the wealth of spectroscopic data only because absorption
is present. A study of absorption effects is given
in LP04. There it was found that for sufficiently thin slices
the scalings obtained in the absence of absorption still hold
provided that the absorption on the scales under study is negligible.
A similar criterion is valid for the VCS.
From the practical point of view, absorption imposes an upper limit
on the scales for which the statistics can be recovered.

If integrated intensity of spectral lines is studied in the presence of
absorption non-trivial effects emerge. Indeed, for optically thin
medium the spectral line integration results in intensity reflecting
the density statistics. LP04 showed that this may not
be any more true for lines affected by absorption. Depending on the
spectral index of velocity and density fluctuations the contributions 
from either from 
density or velocity dominate the integrated intensity fluctuations.
When velocity is dominant a very interesting regime for which
intensity fluctuations show universal behavior, i.e. the
power spectrum $P(K)\sim K^{-3}$  emerges. If density is dominant,
the spectral index of intensity fluctuations is the same
as in the case 
an optically thin cloud. Conditions for these regime as well
as for some more interesting intermediate asymptotic regimes
are outlined in LP04.

An additional point explained in LP04 is that even in the case of
fully integrated spectral lines in the presence of absorption the
velocity effects are still modifying the intensity statistics.

\section{What do observations tell us?}

Application of the VCA to the Galactic data in LP00 and to
Small Magellanic Cloud in Stanimirovic \& Lazarian (2001)
revealed spectra of 3D velocity fluctuations consistent with
the Kolmogorov scaling. LP00 argued that the same scaling
was expected for the magnetized turbulence appealing to
the Goldreich-Shridhar (1994) model. Esquivel et al. (2003)
used simulations of MHD turbulent flows to show that in spite
of the presence of anisotropy caused by magnetic field the
expected scaling of fluctuations is Kolmogorov. Studies by
Cho\& Lazarian (2002, 2003) revealed that the Kolmogorov-type
scaling is also expected in the compressible MHD flows.
These studies support MHD turbulence model for SMC.

\begin{figure} [h!t] 
{\centering \leavevmode 
\epsfxsize=2.5in\epsfbox{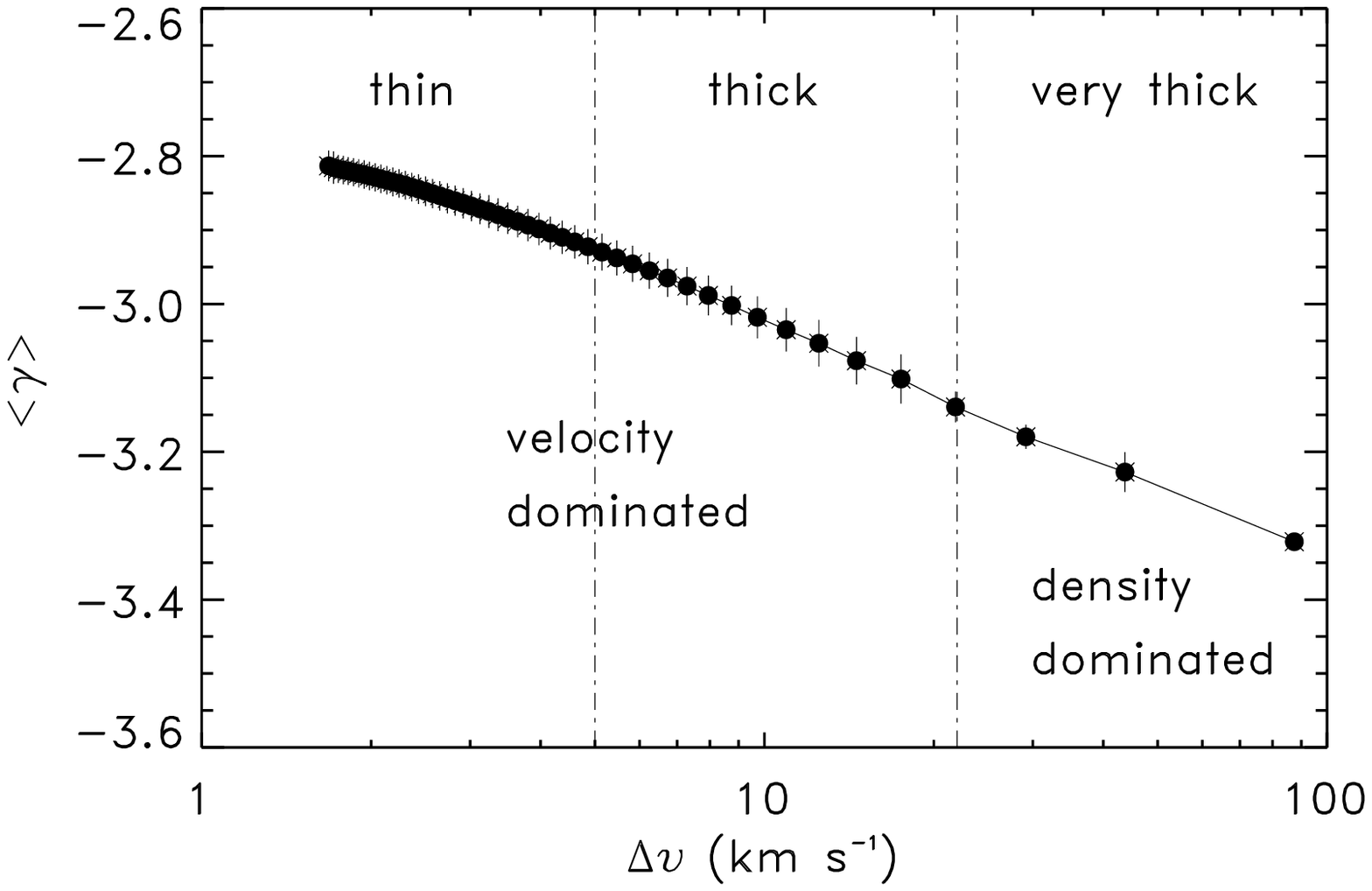} 
\hfil 
\epsfxsize=2.1in\epsfbox{ 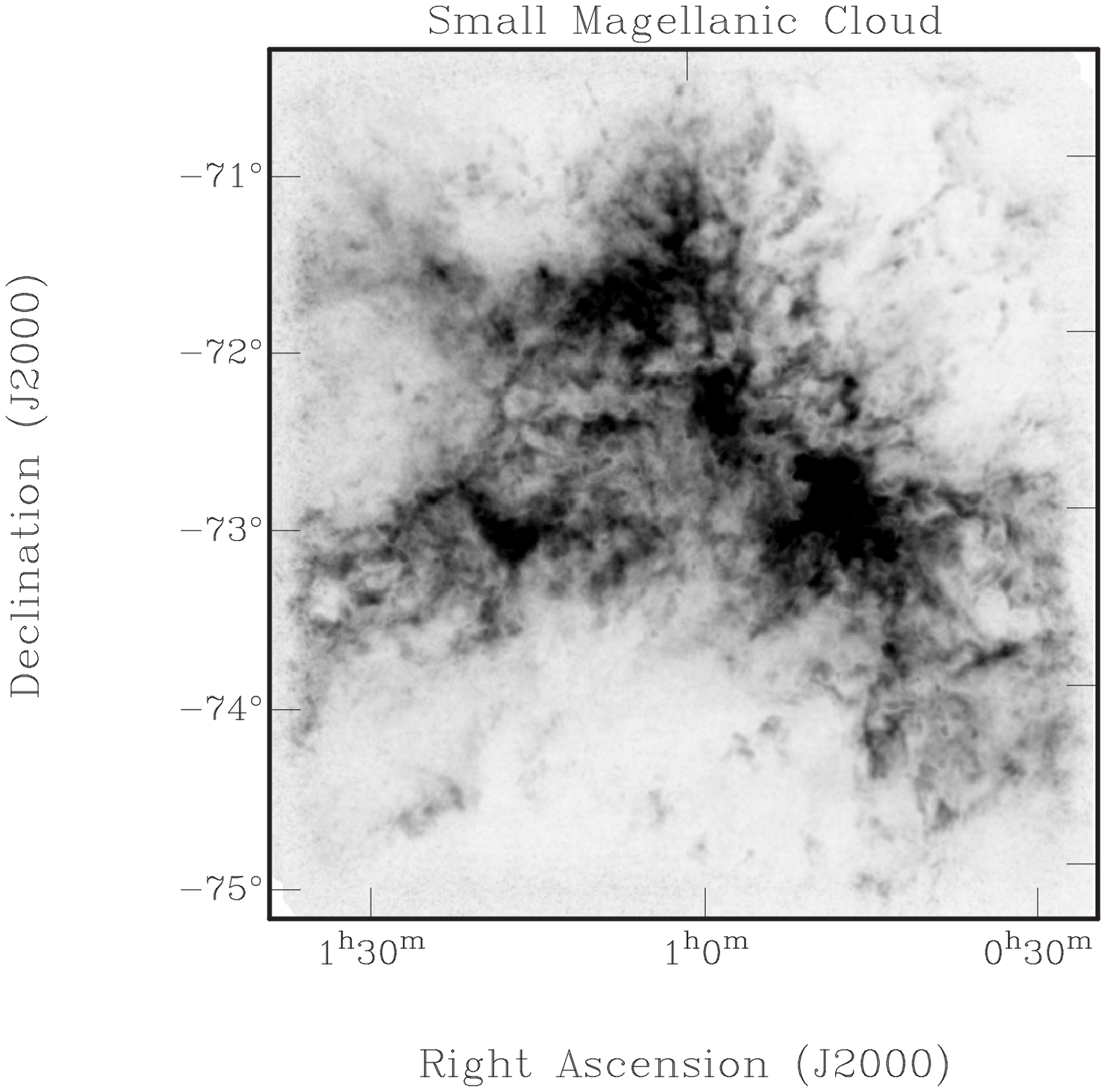} 
} 
\caption{{\it Upper Panel}: Variations of two dimensional 21~cm 
spectral slope with the velocity slice thickness (from Stanimirovic \&
Lazarian 2001). The LP00 study predicts that the thick slice reflects
the density statistics, while the thin slice is influenced by the
velocity, which results in flattening of the slope.
{\it Lower panel}: The 21~cm image of SMC that exhibits a lot
of density structure which, however, provides a subdominant
contribution to thin channel maps.}  
\end{figure}

Studies of turbulence are more complicated for the inner parts of the Galaxy,
where (a) two distinct regions at different distances from the observer
contribute to the emissivity for a given velocity and (b) effects of
the absorption are important. However, the analysis in Dickey et al. (2001)
showed that some progress may be made even in those unfavorable
circumstances. Dickey et al. (2001) found the steepening
 of the spectral index with the increase of the velocity slice thickness.
They also observed the spectral index for strongly absorbing direction
approached $-3$ in accordance with the conclusions in LP04.

21-cm absorption provides another way of probing turbulence on small
scales. The absorption depends on the density to temperature ratio
$\rho/T$, rather than to $\rho$ as in the case of emission. However,
 in terms of the VCA this change is not important and we still expect to
see emissivity index steepening as velocity slice thickness increases,
provided that velocity effects are present. In view of
this, results of Deshpande et al. (2001), who did not see such steepening,
can be interpreted as the evidence of the viscous suppression of
turbulence on the scales less than 1~pc. The fluctuations in this
case should be due to density and their shallow spectrum $\sim k^{-2.8}$ may
be related to the damped magnetic structures below the viscous
cutoff (Cho, Lazarian \& Vishniac 2002b). 

Studies of velocity statistics using velocity centroids are
numerous (see O'Dell 1986, Miesch \& Bally 1994, Miesch, Scalo \& Bally 1999).
The analysis of observational data in Miesch \& Bally (1994) provides a range
of power-law indexes. Their results obtained with structure
functions if translated into spectra are consistent with
$E(k)=k^{\beta}$, where $\beta=-1.86$ with the standard
deviation of $0.3$. The Kolmogorov index $-5/3$
falls into the range of the measured values. L1228 exhibits
exactly the Kolmogorov index $-1.66$ as the mean value,
while other low mass
star forming regions L1551 and HH83 exhibit indexes close
to those of shocks, i.e. $\sim -2$. The giant molecular cloud regions show
shallow indexes in the range of $-1.9<\beta<-1.3$ (see Miesch et al. 1999).
It worth noting that Miesch \& Bally (1994)
obtained somewhat more shallow indexes that are closer to the
Kolmogorov value using autocorrelation functions. Those may
be closer to the truth as in the presence
of absorption in the center of lines, minimizing the regular velocity
used for individual centroids might make the results
more reliable. Whether the criterion given by (\ref{criterion})
is satisfied for the above data is not clear. If it is not
satisfied, which is quite possible as the Mach numbers are
large for molecular clouds, then the spectra above reflect
the density rather than velocity. If the criterion happen to
be satisfied a more careful study
taking absorption effects into account is advantageous. 

Apart from testing of 
the particular
scaling laws, studies of turbulence statistics
 should identify sources and injection
scales of the turbulence.
 Is turbulence in molecular clouds a part
of a large scale ISM cascade (see Armstrong et al. 1997)? 
How does the share of the energy within
compressible versus incompressible motions vary within the Galactic
disk? There are examples of questions that can be answered in
future. 

\section{How to go beyond power spectrum?}

\subsection{\bf Genus Analysis}

Velocity and density power spectra do not provide the complete description
of turbulence. Intermittency of turbulence (its variations in time and space) 
and its topology in the presence of different phases are 
not described by the power spectrum.
 
``Genus analysis" is a good tool for studying the topology of
turbulence (see review by Lazarian 1999), receiving well-deserved
recognition for cosmological studies (Gott et al. 1986). Consider an
area on the sky with contours of projected density. The 2D genus,
$G(\nu)$, is the difference between the number of regions with a
projected density higher than $\nu$ and those with densities lower
than $\nu$. Fig.~3 shows the 2D genus as the function of $\nu$
for a Gaussian distribution of densities (completely symmetric curve),
for MHD isothermal
simulations with Mach number 2.5, and for HI in SMC.
\begin{figure} [h!t] 
{\centering \leavevmode 
\epsfxsize=2.5in\epsfbox{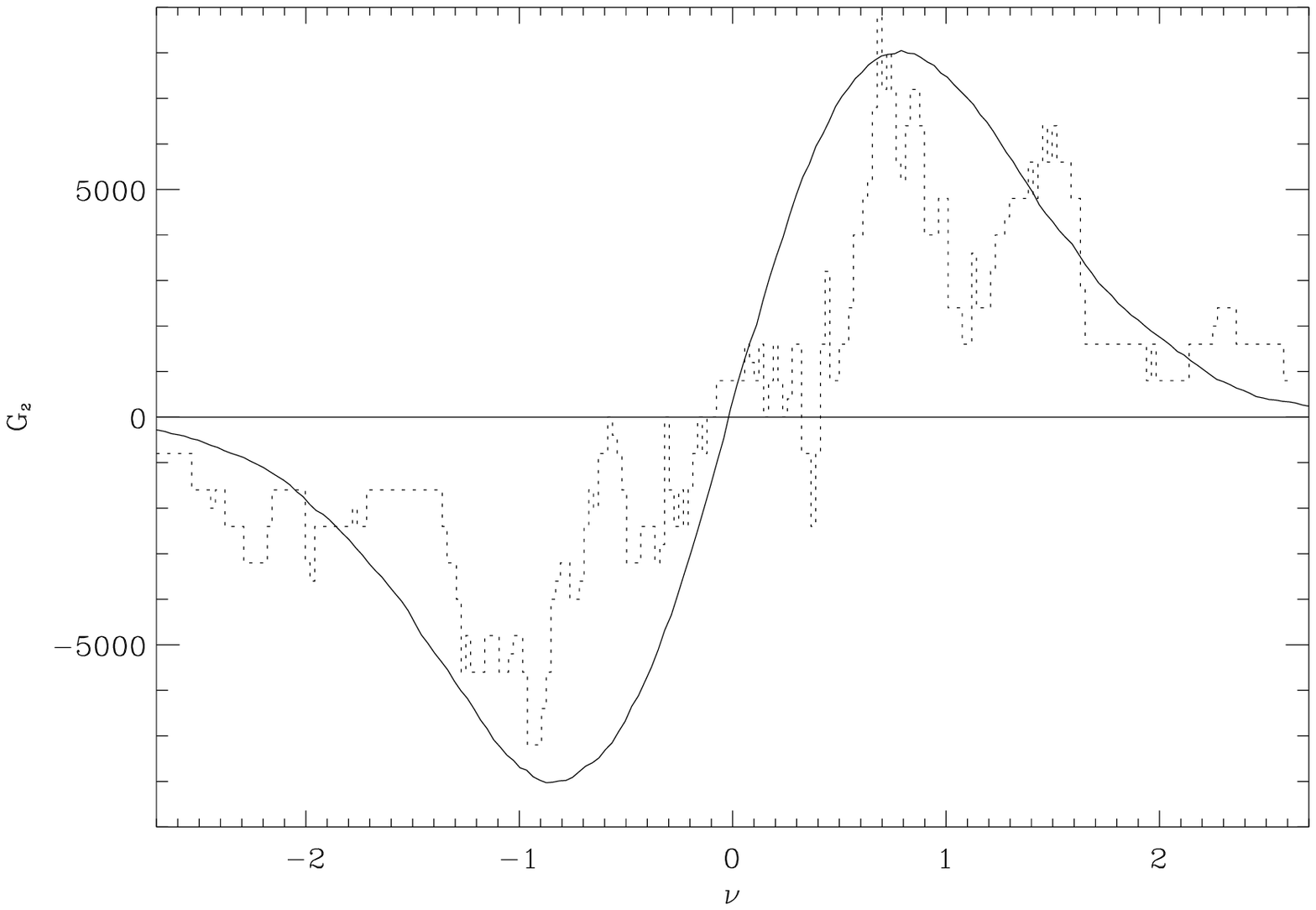} 
\hfil 
\epsfxsize=2.5in\epsfbox{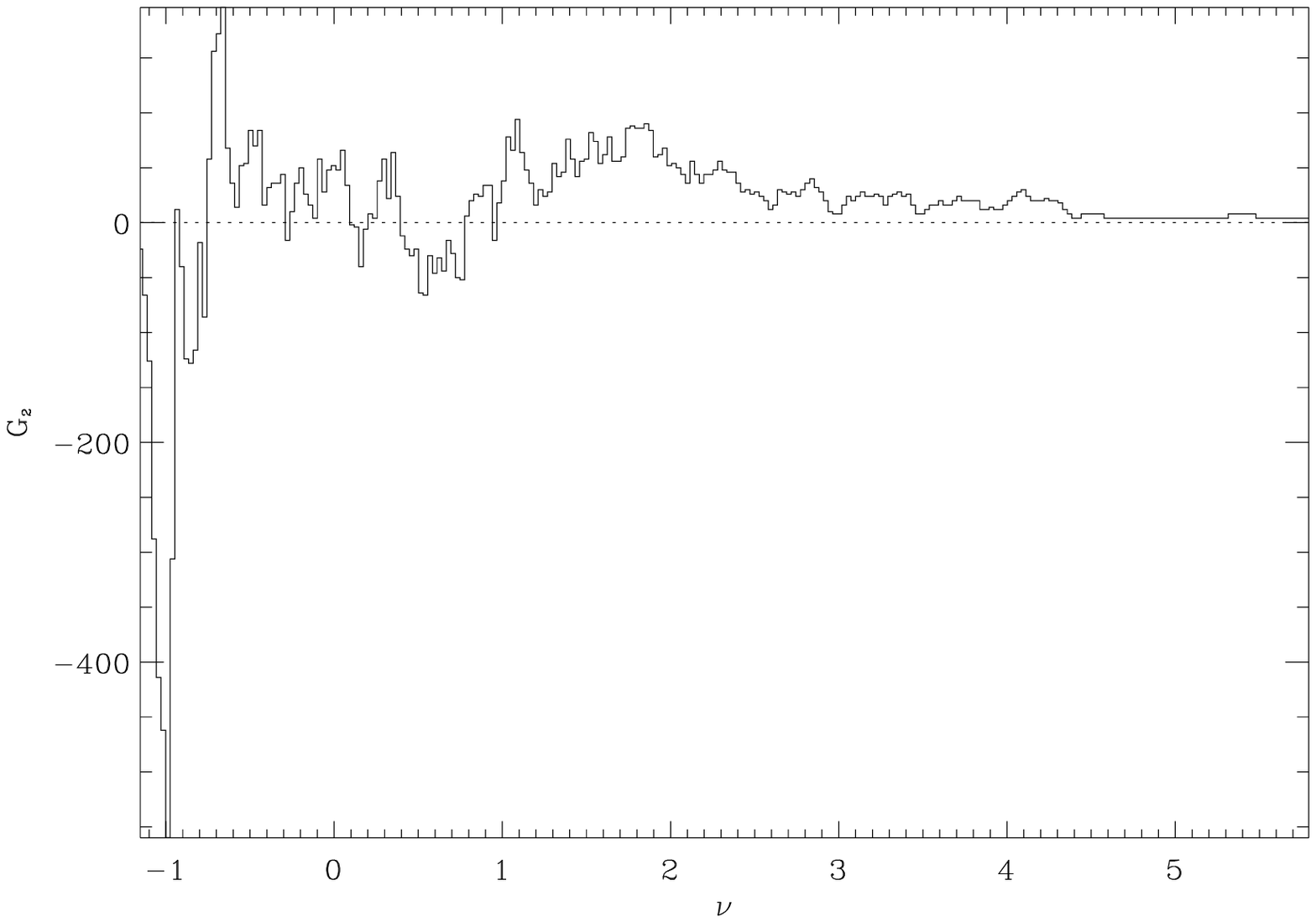} 
} 
\caption{{\it Upper panel} shows the 2D genus of the Gaussian 
distribution 
(smooth analytical curve) 
against the genus for the isothermal compressible MHD simulations with Mach 
number 2.5 (dotted curve). {\it Lower panel} shows the genus of 
HI distribution in SMC. 
The topology of the HI distribution is very different, although the 
spectra are similar (from Esquivel, Lazarian, Pogosyan
\& Cho, in preparation).} 
 \end{figure} 
 
Isothermal MHD simulations exhibit more or less symmetric density
distributions, but the SMC data reflect a prominent Swiss cheese
topology, which also can be suggested from the visual inspection of
the image (see Fig.~3). However, unlike visual inspection, the genus quantifies the
topology and allows us to compare numerical results with
observations. Note, that the MHD simulations in Fig~3 are not so different
from the SMC in terms of the power spectrum.

\subsection{She-Leveque exponents}

The p-th order (longitudinal) velocity structure function $SF_p$ 
and scaling exponents
$\zeta(p)$ are defined as
\begin{equation}
SF_p({\bf r}) \equiv \langle 
  | \left[ {\bf v}({\bf x}+{\bf r})-{\bf v}({\bf x}) \right] \cdot {\bf \hat r} |^p
  \rangle \propto r^{\zeta (p)},
\label{structure}
\end{equation}
where the angle brackets denote averaging over {\bf x}.

The scaling relations suggested by She \& Leveque (1994) 
related $\zeta(p)$ to the scaling of the velocity $v_l\sim l^{1/g}$,
the energy cascade rate $t_l^{-1}\sim l^{-x}$, and the co-dimension of the
dissipative structures $C$:
\begin{equation}
\zeta(p)={p\over g}(1-x)+C\left(1-(1-x/C)^{p/g}\right).
\label{She-Leveque}
\end{equation}
For incompressible turbulence these parameters are $g=3$, $x=2/3$, 
and $C=2$, implying that
dissipation happens over 1D structures (e.g. vortices). Further
studies (see Muller \& Biskamp (2000), Padoan et al. (2003),
Cho, Lazarian \& Vishniac (2003)) has shown that the expected exponents
can be different for the case of MHD especially for the case of the
viscosity-damped regime of turbulence theoretically described in
Lazarian, Vishiniac \& Cho (2004) and numerically studied in
Cho, Lazarian \& Vishniac (2002, 2003), Cho \& Lazarian (2003).

At present the studies of She-Leveque exponents are inconclusive.
Padoan et al. (2003) studied integrated density maps rather than
velocity. While the results were  compared with a particular model
of turbulence dissipation, we would note that the density spectrum
was in itself much shallow than the Kolmogorov one for the case 
studied. The latter is expected to be close to Kolmogorov for
the model they compared their observations with. 
Obtaining the corresponding statistics from observations is potentially possible
using modified centroids. A further study of the issue is absolutely essential.

\subsection{Spectral Correlation Functions}

An interesting tool termed ``spectral correlation function''  or SCF has been
recently introduced to the field (Goodman 1999, Rosolowsky et al. 1999,
Padoan et al. 2001, Balesteros-Peredes et al. 2002, Padoan, Goodman \& Juvela 2003). 
It compares neighboring spectra with each other. For this
purpose the following measure is proposed:
\be
S(T_1, T_0)_{s,l}\equiv 1-\left(\frac{D(T_1,T_2)}{s^2\int T_1^2(v)dv+
\int T_0^2(v)dv}\right)~~~,
\ee
where the function
\be
D(T_1,T_2)_{s,l}\equiv \left\{\int [s T_1(v+l)-T_0(v)]^2dv\right\}
\ee
and the parameters $s$ and $l$ can be adjusted. One way to choose
them is to minimize $D$ function. In this case $S$ function
gets sensitive to similarities in the shape of two profiles. 
Fixing $l$, $s$ or both
parameters one can get another 3 function that are also 
sensitive to similarities in amplitude, velocity offset
or to both parameters.

The purpose of those functions is to distinguish regions with various
star forming activity and to compare numerical models with observations.
To do this histograms of SCF are compared with
histograms of SCF obtained for the randomized spatial positions. 
This allows to models to be distinguished on the basis of their clustering
properties. Results reported by Rosolowsky et al. (1999) are encouraging. 
It was possible to find differences for simulations corresponding to
magnetized and unmagnetized media and for those data sets for which
an earlier analysis by Falgarone et al. (1994) could not find the difference. 
It seems feasible to use wavelets
that will emphasize some characteristics of the histograms in order
to make the distinction quantitative. Construction these wavelets
will be the way of ``teaching'' SCF to  extract 
features that distinguish various sets of data.

A few comments about spectral correlation functions may be relevant.
First of all, by its definition it is a flexible tool. In the
analysis of Rosolowsky et al. (1999) the SCF were calculated for the
subcubes over which the original data was divided. In this way SCF
preserves the spatial information and in some sense is similar
to cloud-finding algorithms (see Stutzki \& Gusten 1990, Williams,
de Geus \& Blitz 1994). If, however, one fixes the angular separation between
the studied spectra and then the technique will be more similar to
the traditional correlation function analysis that is sensitive to
turbulence scale rather than to positional information. If
SCF are applied for the velocity slices they should reproduce
the results of VCA (see LP00). The application of the SCF to the data along
the velocity coordinate is expected to give a universal $V^{-2}$,
 as the power spectra along
the velocity coordinate is too steep for correlation functions
to reflect it (LP04). This automatically indicates that much care should
be used when interpreting SCF that includes $V$-separations.

\subsection{Anisotropy Analysis}
In isotropic turbulence, correlations depend only on the distance
between the points. Contours of equal correlation are 
circular in this case. Presence of magnetic field introduces anisotropy
and these contours become elongated with a symmetry axis given by
the magnetic field.  To study turbulence
anisotropy, we can measure contours of equal correlation corresponding
to the data within various
velocity bins. The results obtained with simulated data are shown in Fig.~3. 
\begin{figure} [h!t] 
{\centering \leavevmode 
\epsfxsize=2.5in\epsfbox{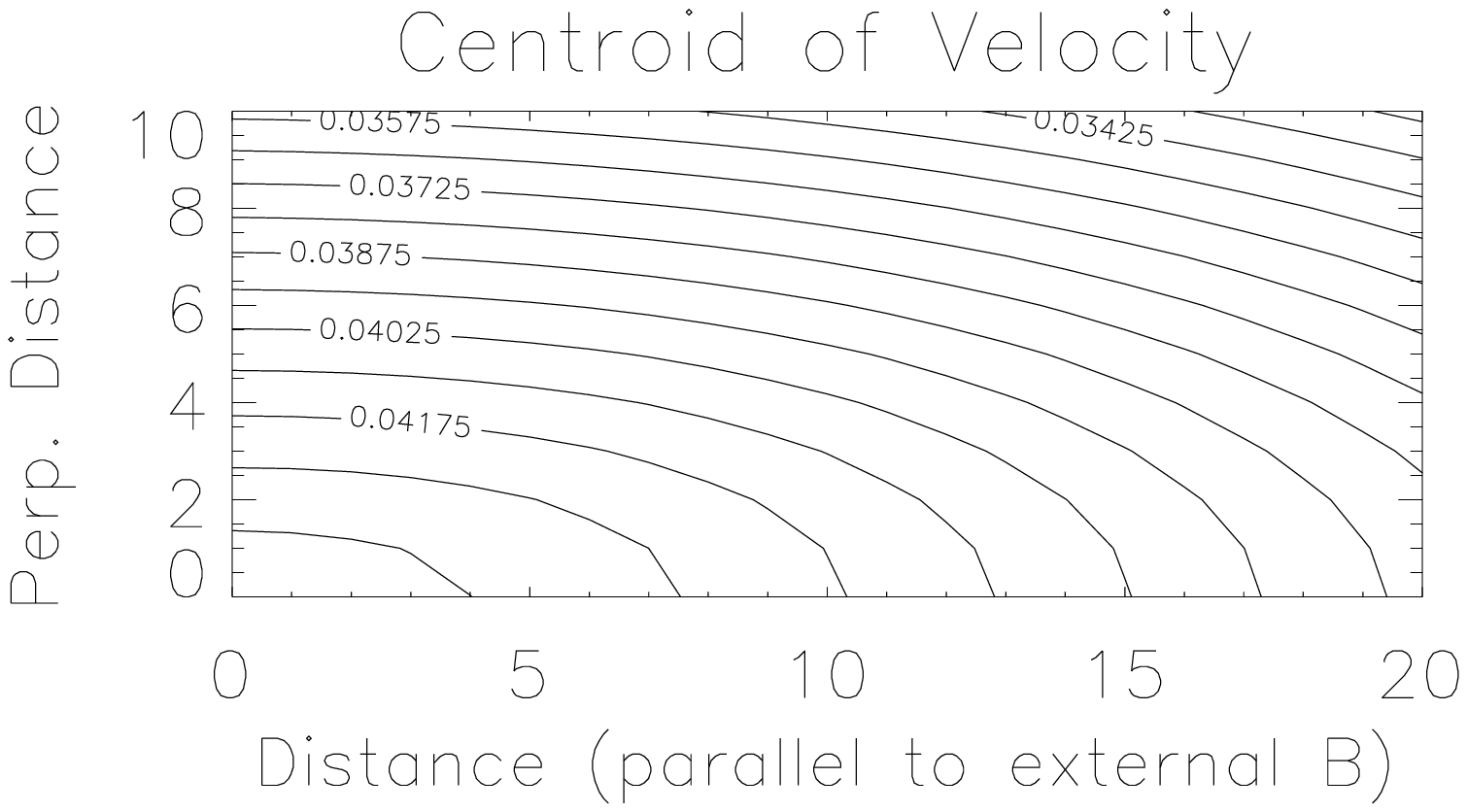} 
\hfil 
\epsfxsize=2.5in\epsfbox{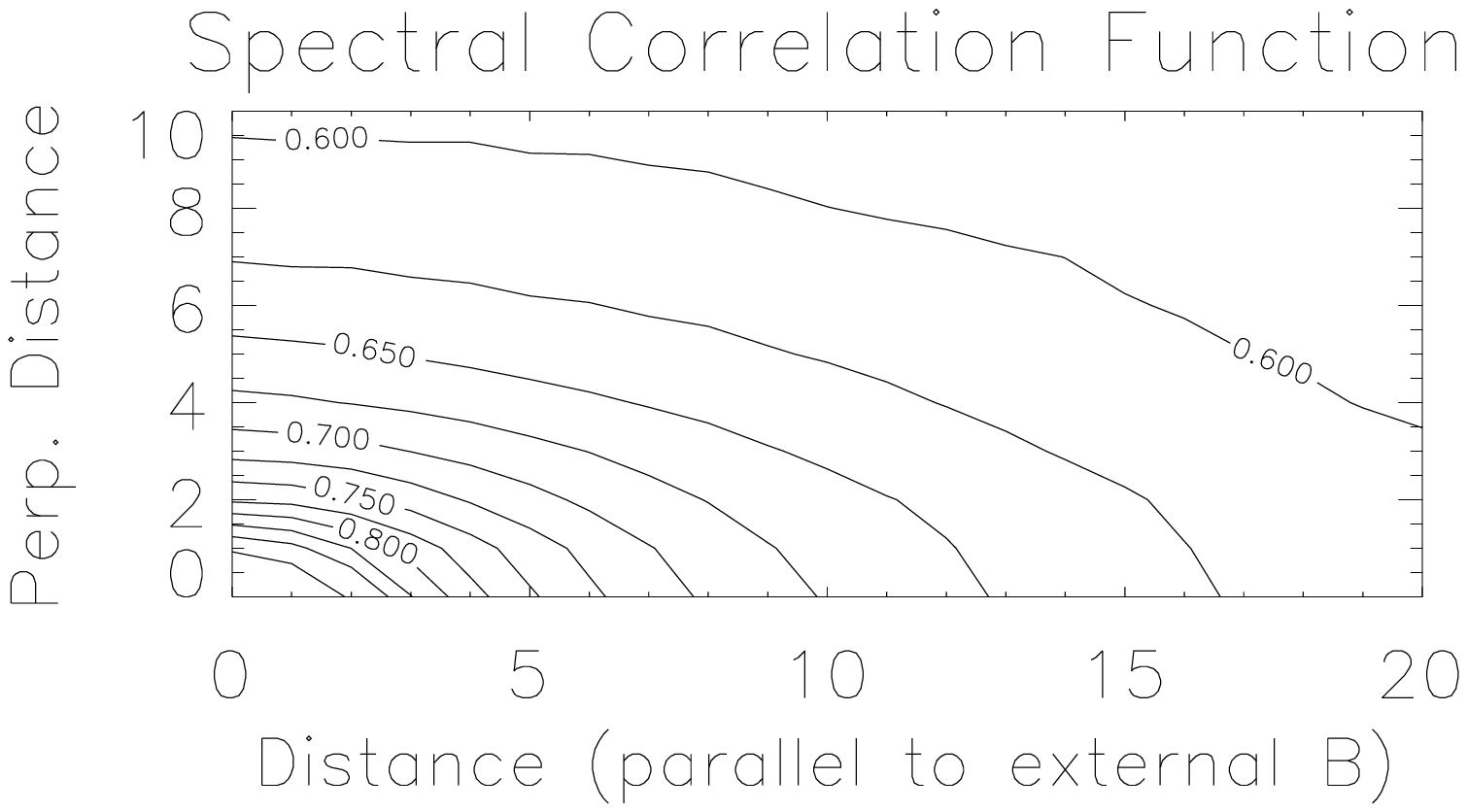} 
}  
\caption{{\it Anisotropies of the synthetic data by Cho, Esquivel} \&
{\it Lazarian}. Anisotropies measured by correlations of
Centroids of Velocity ({\it left})
and by Spectral Correlation Function (SCF)({\it right}) reveal the direction of
magnetic field. Combined with the anisotropy analysis, the SCF (introduced by 
Alyssa Goodman) is likely to become even more useful tool.}  
\end{figure}

Since the anisotropy is related
to the magnetic field, the studies of anisotropy can provide means to analyze
magnetic fields. It is important to study different data sets and channel
maps for the anisotropy. Optical and infrared polarimetry
can benchmark the anisotropies in correlation
functions. We hope that the
anisotropies will reveal magnetic field within dark clouds where grain
alignment and therefore polarimetry fails (see Lazarian \& Cho 2005,
for a review of grain alignment).

\section{Summary}

1. A criterion exist that determines whether Velocity Centroids provide a good measure
of turbulent velocity. The performance of Velocity Centroids can be improved via
using Modified Velocity Centroids, that have a reduced contribution from density
fluctuations. Numerical studies show that Velocity Centroids are most reliable at
low Mach numbers.

2. Velocity Channel Analysis (VCA) has been tested to be a reliable tool for studies of
supersonic interstellar turbulence. Results obtained via VCA reveal Kolmogorov-type
turbulence in Milky Way and Small Magellanic Cloud galaxy.

3. Velocity Coordinate Spectrum (VCS) is a new tool for studies supersonic turbulence.  
It allows to study turbulence even when the spatial resolution is not adequate.
This provides a unique way for studies extragalactic turbulence. 

4. Absorption modifies results of VCA and VCS. It places upper limits on the scales
that can be studied by either of  the techniques. It introduces more mixing of velocity
and density that results in the statistics of
 integrated line intensity being not sensitive to underlying
density. Nevertheless studies of  small scale turbulence are possible with sufficiently
thin velocity slices of data.

5. While power spectra is a useful measure of turbulence, it is useful to go beyond
it. Studies of topology, anisotropy of turbulence is possible with other tools that
we briefly discuss.

\acknowledgements{This work is supported by the Center for Magnetic Self-Organization
in Laboratory and Astrophysical Plasmas. Fruitful discussions with
Alexey Chepurnov, Alejandro Esquivel, Shusaku Horibe and Dmitry Pogosyan are acknowledged}

\end{document}